\documentclass[fleqn,usenatbib]{mnras}

\usepackage{newtxtext,newtxmath}

\usepackage[T1]{fontenc}
\usepackage{ae,aecompl}

\usepackage{graphicx}	
\usepackage{amsmath}	
\usepackage[T1]{fontenc}
\usepackage[utf8]{inputenc}
\usepackage{xcolor}

\usepackage{hyphenat}
\title[S-PLUS: MSPs in GCs]{S-PLUS: Exploring wide field properties of multiple populations in galactic globular clusters at different metallicities}

\author[E. A. Hartmann et al.]{
Eduardo A. Hartmann$^{1}$\thanks{E-mail: eduardo.hartmann@ufrgs.br (UFRGS)},
Charles J. Bonatto$^{1}$\thanks{E-mail: charles@if.ufrgs.br},
Ana L. Chies-Santos$^{1,2}$\thanks{E-mail: ana.chies@ufrgs.br},
Javier \newauthor{Alonso-Garc\'ia$^{3,4}$, 
Nate Bastian$^{5,6,7}$, 
Roderik Overzier$^{9,8}$,
William Schoenell$^{10}$,
Paula} \newauthor{R. T. Coelho$^{8}$,
Vinicius Branco$^{8}$,
Antonio Kanaan$^{11}$,
Claudia Mendes de Oliveira$^{8}$}, 
\newauthor{Tiago Ribeiro$^{12}$}
\\
$^{1}$Departamento de Astronomia, Instituto de F\'isica, UFRGS, Av. Bento Gonçalves, 9500, Porto Alegre, RS, Brazil\\
$^{2}$Shanghai Astronomical Observatory, Chinese Academy of Sciences, 80 Nandan Rd 200030, Shanghai, China\\
$^{3}$Centro de Astronom\'{i}a (CITEVA), Universidad de Antofagasta, Av. Angamos 601, Antofagasta, Chile\\
$^{4}$Millennium Institute of Astrophysics, Nuncio Monse\~nor Sotero Sanz 100, Of. 104, Providencia, Santiago, Chile\\
$^{5}$Donostia International Physics Center (DIPC), Paseo Manuel de Lardizabal, 4, 20018, Donostia-San Sebastián, Guipuzkoa, Spain\\
$^{6}$IKERBASQUE, Basque Foundation for Science, 48013, Bilbao, Spain\\
$^{7}$Astrophysics Research Institute, Liverpool John Moores University, Liverpool, L3 5RF,United Kingdom \\
$^{8}$Universidade de S\~ao Paulo, S\~ao Paulo, Instituto de Astronomia, Geof\'isica e Ci\^encias Atmosf\'ericas, SP, Brazil\\
$^{9}$Observat\'orio Nacional, Rua General Jos\'e Cristino, 77, S\~ao Crist\'ov\~ao, 20921-400, Rio de Janeiro, RJ, Brazil\\
$^{10}$GMTO Corporation 465 N. Halstead Street, Suite 250 Pasadena, CA 91107\\
$^{11}$Departamento de F\'isica, Universidade Federal de Santa Catarina, Florian\'opolis, SC, 88040-900, Brazil\\
$^{12}$NOAO, P.O. Box 26732, Tucson, AZ 85726\\
}

\date{Accepted XXX. Received YYY; in original form ZZZ}

\pubyear{2022}

\begin{document}
\label{firstpage}
\pagerange{\pageref{firstpage}--\pageref{lastpage}}
\maketitle

\begin{abstract}
Multiple Stellar Populations (MSPs) are a ubiquitous phenomenon in Galactic Globular Clusters (GCs). By probing different spectral ranges affected by different absorption lines using the multi-band photometric survey S-PLUS, we study four GCs ---NGC104, NGC288, NGC3201 and NGC7089--- that span a wide range of metallicities. With the combination of broad and narrow-band photometry in 12 different ﬁlters from 3485A (u) to 9114A (z), we identiﬁed MSPs along the rectiﬁed red-giant branch in colour-magnitude diagrams (CMDs) and separated them using a K-means clustering algorithm. Additionally, we take advantage of the large Field of View of the S-PLUS detector to investigate radial trends in our sample. We report on six colour combinations that can be used to successfully identify two stellar populations in all studied clusters and show that they can be characterised as Na-rich and Na-poor. For both NGC 288 and NGC 7089, their radial proﬁles show a clear concentration of 2P population. This directly supports the formation theories that propose an enrichment of the intra-cluster medium and subsequent star formation in the more dense central regions. However, in the case of NGC 3201, the trend is reversed. The 1P is more centrally concentrated, in direct contradiction with previous literature studies. NGC 104 shows a well-mixed population. We also constructed radial proﬁles up to 1 half-light radius of the clusters with HST data to highlight that radial differences are lost in the inner regions of the GCs and that wide-ﬁeld studies are essential when studying this.

\end{abstract}




\section{Introduction}
\label{S:Intro}

\begin{figure*}
    \centering
    \includegraphics[width=0.95\textwidth]{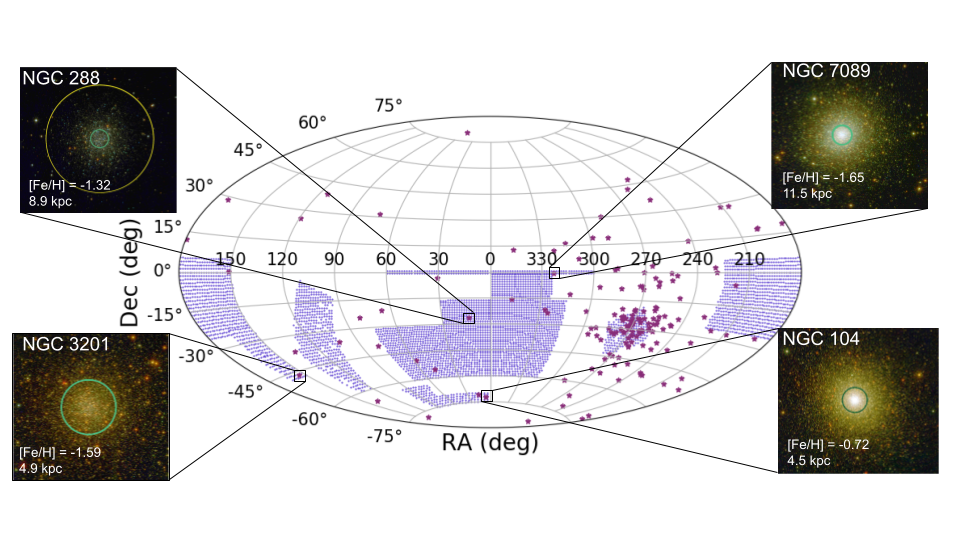}
    \caption{All sky view of the S-PLUS survey footprint in blue. The purple stars are the GCs of our Galaxy from \citet{Harris1996}, 2010 version. The insets show the S-PLUS FoV of the four clusters in our sample, as well as their metalicities and heliocentric distances. In these insets, we have highlighted by green circles the core radius and in yellow the tidal radius.}
    \label{fig:aitoff}
\end{figure*}

\par The phenomenon of multiple stellar populations (MSPs) in Galactic globular clusters (GCs) has been well observed in almost all GCs older than 2 Gyr, both spectroscopically and photometrically \citep{Piotto2015}. Detailed spectroscopical studies of stars in GCs have shown that there are significant abundance variations in light elements not compatible with a single stellar population \citep{Carretta2009}. Increased abundances of He, N and Na with a decrease in C and O are the telltale sign of the Second Population (2P) of stars in a cluster. Stars without these characteristics are considered the First Population (1P).

To explain such abundance variations, some formation scenarios propose that the first population of stars enriches the intra-cluster medium forming a second population (or more, e.g. NGC 7089 with 7 populations). This scenarios differ primarily on what pollutes the next generation, be that Asymptotic Giant Branch (AGB) stars \citep{D'ercole2016}, fast-rotating massive stars (FRMS) \citep{Decressin2007,Charbonnel2014}, interacting binary \citep{deMink2009} or a supermassive star \citep{Gieles2018} (SMS). One of the big challenges is the mass-budget problem; in short, the observed 2P in clusters is larger (or at least equal to) in number than the 1P. In the popular enrichment scenarios, there would not be enough processed material by the first population for the formation of the second \citep{Prantzos2006}. Possible solutions for this include the assumption that clusters were ~30 -- 60 times more massive than at present-day and that 90 -- 95\% of 1P stars were lost during the evolution of these objects. \citet{Wang2020} developed a scenario based on stellar mergers of binaries when the cluster is very young. This has the advantage of combining previous scenarios such as the FRMS and SMS with merging binaries and introduces the necessary stochasticity observed in GCs. Their results suggest that this may play an important role in forming MSPs, avoiding such pitfalls as the mass-budget problem.

\par In the most common proposed explanations for MSPs, the 2P is formed in the more dense central region of the cluster. One of the ways to test this is by constructing cumulative radial distributions of both populations. \citet{Lardo2011} looked at the radial profile of nine GCs and concluded that the enriched population is more centrally concentrated in the majority of the clusters. Nevertheless, \citet{Hoogendam2021} reanalysed the data and found that the populations are not as segregated as was thought, recommending caution when doing such studies. \citet{Dalessandro2019} studied 20 GCs of various dynamical ages using the parameter A$^{+}$ to quantify the difference between both populations. They showed that the second population is more centrally concentrated in dynamically younger clusters, while no significant difference exists in older GCs. This directly supports the idea that the 2P was formed more centrally concentrated in the cluster.

\par While spectroscopic studies are limited to small samples of very bright stars in clusters that may contain up to millions of them, high precision multi-band photometric studies have been able to identify and characterise the different populations found in GCs with the advantage that they are able to analyse thousands of stars simultaneously \citep[e.g.,][]{Lardo2011, Soto2017, Larsen2019}. The spread in metallicity manifests itself in colour-magnitude diagrams (CMD) as different evolutionary sequences of stars when appropriate filters are used, especially those that capture conspicuous metallic features. The Hubble UV Legacy Survey has been incredibly successful in using the filters of the Hubble Space Telescope (HST) to separate populations \citep{Piotto2015}. In addition, pseudo-colours (the difference between two colours) has been a great tool to separate the populations better \citep{Milone1}.

\begin{table*}
\caption{Characteristics of the studied GCs, heliocentric distance, metallicity, interstellar reddening, half-mass radius and tidal radius. All information from \citet{Harris1996}, 2010 version.}
\begin{tabular}{l|c|c|c|c|c}
Name      & Distance & {[}Fe/H{]} & E(B-V) & r$_{h}$  & r$_{t}$   \\ 
          & (kpc)    &            & (mag)  & (arcmin) & (arcmin)  \\ \hline
NGC\,104  & 4.5      & -0.72      & 0.04   & 3.17     & 42.86     \\
NGC\,288  & 8.9      & -1.32      & 0.03   & 2.23     & 12.94     \\
NGC\,3201 & 4.9      & -1.59      & 0.24   & 3.1      & 28.45     \\
NGC\,7089 & 11.5     & -1.65      & 0.06   & 1.06     & 21.45           
\end{tabular}
\label{tab:GC_info}
\end{table*}


\par The Southern Photometric Local Universe Survey (S-PLUS; \citealt{SPLUS0}) is observing a considerable area of the southern sky in 12 optical filters, including a fraction of the Milky Way GC system, as can be seen in Fig.~\ref{fig:aitoff}. \citet{CharlesM15} have studied the Cluster M\,15 using observations from the Javalambre Photometric Local Universe Survey (J-PLUS; \citealt{Cenarro2019}), which uses the same filter system and instrument as S-PLUS. They have shown that the combination of blue and red filters can separate the two sequences of stars in the top of the Red Giant Branch (RGB). It is important to note that this cluster is very metal-poor ([Fe/H] $\sim$ $-2.3$) and is located at $10.4$\,kpc from the Sun. In the aforementioned paper, they show two synthetic spectra from \citet{Coelho2011} and \citet{Coelho2014}, one with a primordial composition and another with an enhanced metallicity. Although only qualitatively, this helps elucidate the origin of the splits seen in the CMDs. These results are promising, and they suggest studying more metal-rich clusters through the filters of J-PLUS/S-PLUS.

\par Following this analysis, we have chosen four Galactic GCs in the S-PLUS footprint for this study. They are NGC\,104, NGC\,288, NGC\,3201 and NGC\,7089 and encompass one dex in metallicity, from NGC\,7089 with [Fe/H]\,$=\,-1.65$ to NGC\,104 with [Fe/H]\,$=\,-0.72$. Moreover, they are relatively close by (see Tab.\,\ref{tab:GC_info}), which makes them ideal candidates to explore the phenomenon of MSPs in different environments and varied intrinsic properties.

\par This paper is organised as follows: in section \ref{S:Methodology} we detail the extraction and calibration of the S-PLUS photometry in the cluster fields and the identification of sources as cluster members. In section \ref{S:Sub-pops} we explore the separation of the populations, and in section \ref{S:Analyses} we present the analysis of the clusters and their populations. Finally, in section \ref{S:Conclusion}, we present our conclusions.


\section{Methodology}
\label{S:Methodology}

S-PLUS \citep{SPLUS0} is a photometric survey that is observing $\sim$9300 deg$^2$ of the southern sky in twelve filters: five broad bands (u, g, r, i, z) and seven narrow-bands (F0378, F0395, F0410, F0430, F0515, F0660, F0861). These bands are a subset of the Javalambre filter system and have been chosen for their success in identifying key spectral features in galaxies and stars. It uses the T80-South, an 0.8m telescope in a German equatorial mount located at Cerro Tololo Inter-American Observatory. The detector has a size of $9232 \times 9216$ pixels with a plate scale of $0.55''$\,pixel$^{-1}$ and a Field of View (FoV) of $1.4 \times 1.4$ deg$^2$. Data release two (DR2) \citep{SPLUS2} of S-PLUS covers 950 deg$^2$ with an updated calibration as well as value-added catalogues containing photometric redshift and star-galaxy classification. Since the survey covers the southern part of the sky where the majority of Galactic GCs are, many have been or will be observed by S-PLUS. However, the data reduction pipeline uses {\sc SExtractor}, which is not ideal for identifying sources in crowded fields, such as the central regions of GCs. Thus, to study these objects, we performed Point Spread Function (PSF) photometry using the DAOPHOT \citep{Stetson1987} package in the Image Reduction and Analysis Facility \citep[IRAF;][]{Tody1993} in all 12 filters for the four GCs in our sample. Fig.~\ref{fig:aitoff} shows the area of the sky being observed by S-PLUS and the Milky Way GCs from \citet[ 2010 version]{Harris1996}. In the insets, we highlight the four GCs studied in this work with their colour images constructed using the Trilogy code \citep{Coe2012}.

\subsection{Field-Cluster Star Separation}
\label{S:selection}

\begin{figure*}
    \centering
    \includegraphics[width=0.9\textwidth]{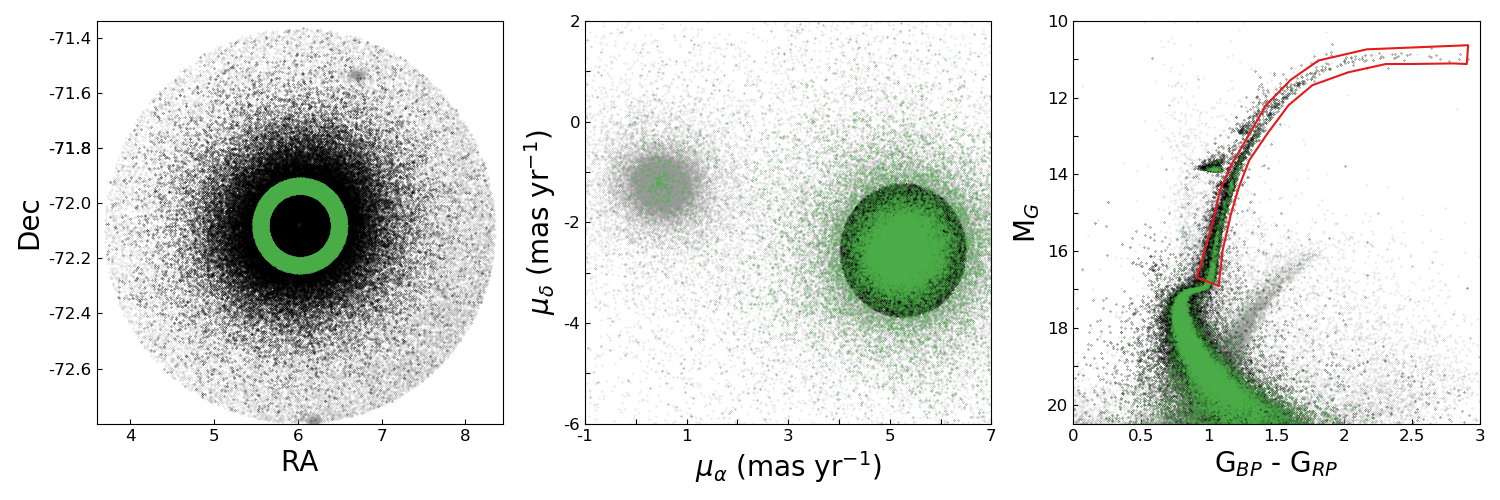}
    \caption{Example of GAIA selection of cluster members for NGC\,104. The three panels share the same colour scheme: in grey are all the objects in the field, in green are the selected representative cluster stars, and in black are the GC members. The first panel shows their spatial distribution in right ascension and declination. The second panel is the proper motion space where the cluster locus is very apparent, and the third panel is the CMD using GAIA colours as well as the RGB polygon. This process is applied to all GCs in the sample.}
    \label{fig:GAIA_selection}
\end{figure*}

Given the large FoV ($\sim$2 deg$^2$) of S-PLUS and our intention of studying the clusters up to their tidal radii, a robust selection process has to be implemented in order to eliminate the maximum number of contaminant objects. For this, the early Data Release 3 of GAIA \citep{GAIA_EDR3} provides us with high precision proper motion information on millions of stars, and we can use them to separate cluster members with a high confidence level. To this end, we first select all objects within the tidal radius (taken from \citealt{Harris1996}, 2010 version) of each cluster and eliminate the ones with one or more unavailable magnitudes and with proper motion errors larger than 1.5 mas yr$^{-1}$. We selected a ring around the centre of the cluster to contain a representative population of the cluster. This way, the GC proper motion locus can be easily identified, and we avoid any crowding issues in the cluster centre. We proceed by fitting a 2D Gaussian profile proper motion space to find the average proper motion of the cluster stars within the ring. We define as cluster members all objects that are inside an ellipse of 5$\sigma$ around the centre of the proper motion distribution. This process is illustrated for NGC\,104 in Fig.~\ref{fig:GAIA_selection}, where we also show a GAIA CMD. In our subsequent analysis, we will concentrate only on the RGB stars. We identify stars in this CMD region through a visually defined polygon, as shown in the right panel in Fig.~\ref{fig:GAIA_selection}. In Appendix \ref{app:selection} we show the same figure for the other three GCs.

\subsection{Differential Calibration}
\label{S:dif_cal}

Zero points (ZPs) for our GC fields were not available at the start of this work. Therefore, we employ the following methodology to calibrate our photometry. We use the fact that in the studied regions, there is a GC that can be well represented by an isochrone and use the code fitCMD (\citealt{fitCMD}) with HST archival data \citep{Piotto2015} in order to obtain the best parameters of each cluster, such as age, metallicity, distance modulus and reddening. The code simulates a population of stars using an initial mass function (IMF) and searches for the best parameters to represent the real population in the CMD. With this information, we obtain the PARSEC isochrones \citep{Bressan2012} with S-PLUS magnitudes for each cluster. The next step consists of 'fitting' the instrumental magnitudes to the corrected isochrone. This is done by first guessing by eye the values for the ZPs, then using a python code that explores the parameter space around the initial guess. It minimises a fitness function that is defined as the sum of the distance from each point to the isochrone in the CMD plane. This is achieved using the simulated annealing method \citep{Kirkpatrick1983}. This process is performed simultaneously in many CMD planes with different colour combinations (e.g. for filters g, r and i, the combinations are (g - r), (g - i) and (r - i), resulting in 9 CMDs). We note that first, we attempted to obtain the ZPs by only applying a correction to the magnitude; however, sometimes this is not enough, and a minor correction must be applied to the colours. This process can only yield good results if the stars used are highly likely members of the cluster and can be represented by the isochrone. For this, we utilise the sample selected through GAIA as described in the previous section. 

\subsection{Differential reddening in NGC3201}
\label{S:dif_red}

\begin{figure}
    \centering
    \includegraphics[width=0.5\textwidth]{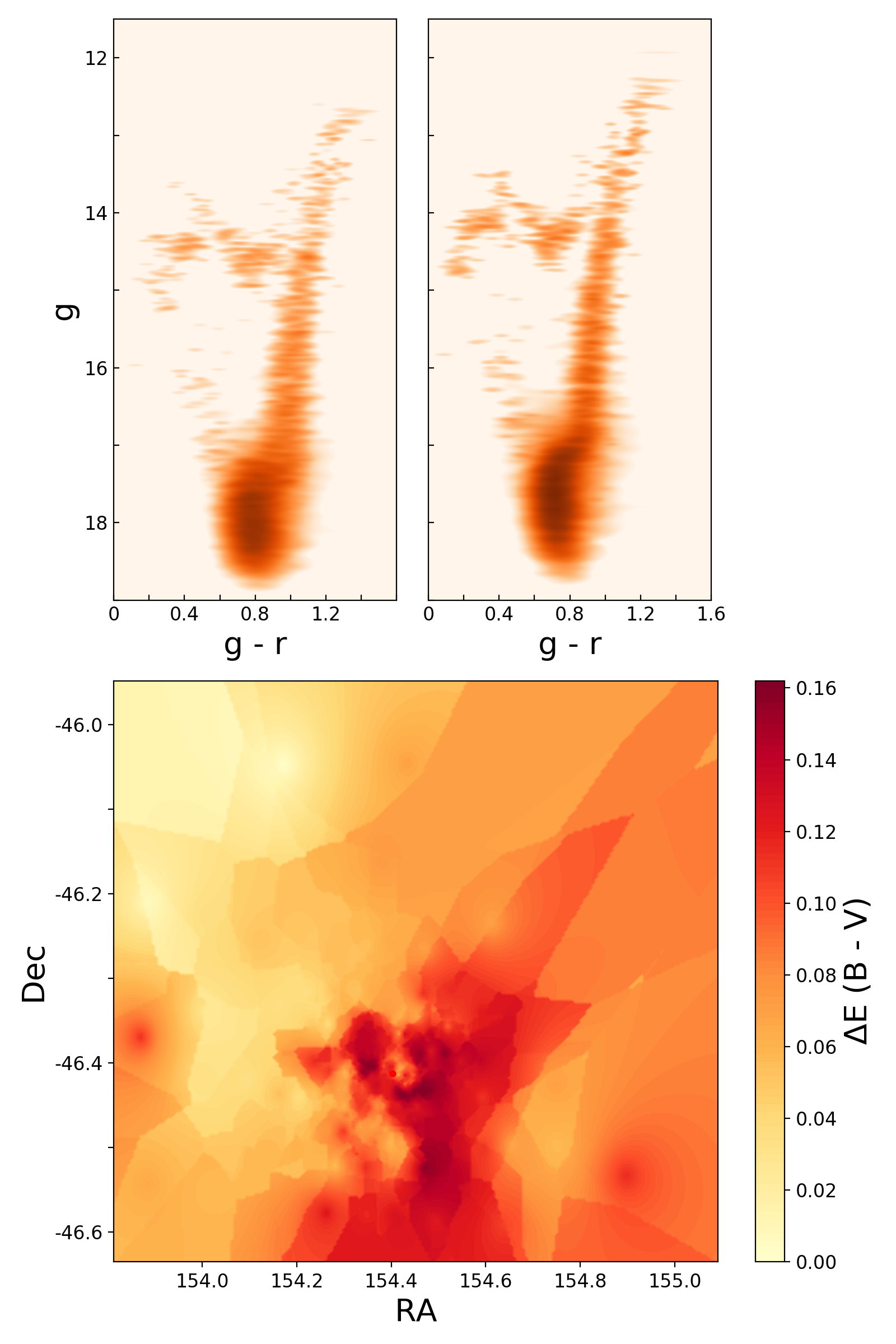}
    \caption{Illustration of differential reddening correction for NGC\,3201. On the left are the original values for the cluster members, and on the right, the CMD is corrected by differential reddening following \citet{Bonatto+Chies2020}. We can see a much more defined turn-off point as well as a narrower RGB. The bottom panel shows the differential reddening map.}
    \label{fig:Diff_Red}
\end{figure}

An additional problem in the analysis of NGC\,3201 is the presence of significant differential reddening. While this can be safely ignored in the other clusters of our sample given their low reddening (see Tab. \ref{tab:GC_info}) and distance from the galactic plane, the effects in NGC\,3201 are more significant (see left panel of Fig.~\ref{fig:Diff_Red}). To correct for this effect, we have used the code described in  \cite{Bonatto+Chies2020}. Briefly, the cluster is divided into cells of at least 50 objects, the CMDs for each cell is constructed, and the bluest one is taken as reference. All others are then shifted to match the reference one, and a reddening value for each sub-region is found. It is important for this analysis that the filter combination used is not affected by the presence of MSPs. For this, we have selected (g - r). Figure \ref{fig:Diff_Red} shows two CMDs of NGC\,3201, before and after the correction and the differential reddening map on the bottom panel. We have also compared the reddening map we obtained with the ones available in the literature, such as \cite{VonBraun2001} and found them to be in agreement both in general shape and values. To test our assumption that the other 3 GCs do not have significant differential reddening we have applied the same code and found that the maximum $\Delta$E\,(B - V) does not exceed ~$0.06$.


\section{The Sub-Populations of Stars}
\label{S:Sub-pops}

\begin{figure}
    \centering
    \includegraphics[width=0.45\textwidth]{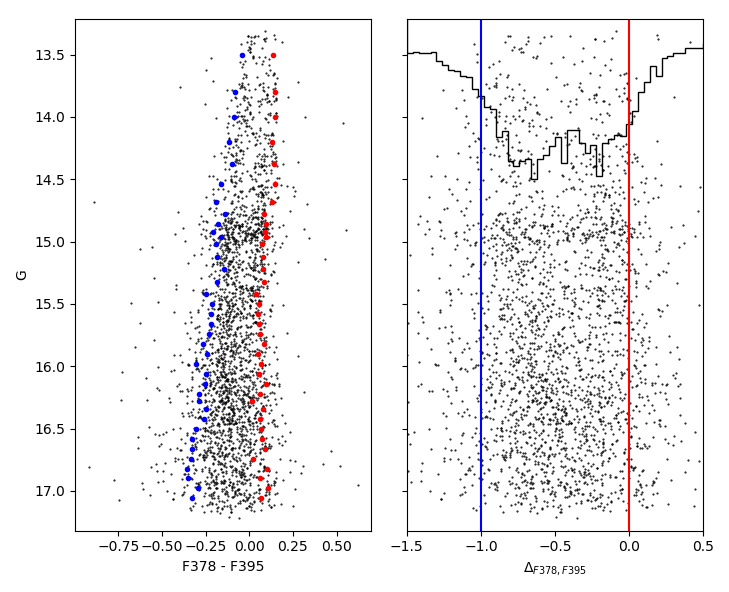}
    \caption{Illustration of the process of making $\Delta_{F378,F395}$ for NGC\,104. The left panel is a CMD showing only the limited RGB. The red and blue dots show the 10\% and 90\% percentiles of each horizontal strip. On the right are the straightened lines and the calculated $\Delta$, as well as a histogram of the horizontal axis showing the double-peaked distribution. The same process is done for all other colours in the clusters in the sample.}
    \label{fig:Make_Delta}
\end{figure}

In order to study the multiple populations, we need to separate them. First, we conduct a visual inspection of all colour combinations in search of a broadening of the RGB, an indication of the presence of MSPs. Six colour combinations are selected, namely: u - F378, u - F395, F378 - F395, F378 - F430, F378 - F515 and F410 - F430. With this, we then constructed the ${\Delta}colours$ in the same manner as \cite{Milone1}. To summarise the process illustrated in Fig.~\ref{fig:Make_Delta}: the RGB is divided in vertical segments containing a minimum of 50 stars, in each segment the 10\% and 90\% percentiles horizontally are determined and are used to create two fiducial lines, the red and blue lines shown in the left panel of Fig.~\ref{fig:Make_Delta}. Such lines are then used to straighten the RGB and create the $\Delta$'s using the same expression from \citet[section 4]{Milone1} as seen in the right panel of Fig.~\ref{fig:Make_Delta}. This process allows the straightening of the RGB and provides a clearer separation of the two populations. This process is repeated for all six colours in all clusters of our sample. One important caveat is that the top of the RGB is ignored in this process because it has a low number of stars, making the process not statistically significant.

\begin{figure*}
    \centering
    \includegraphics[width=0.95\textwidth]{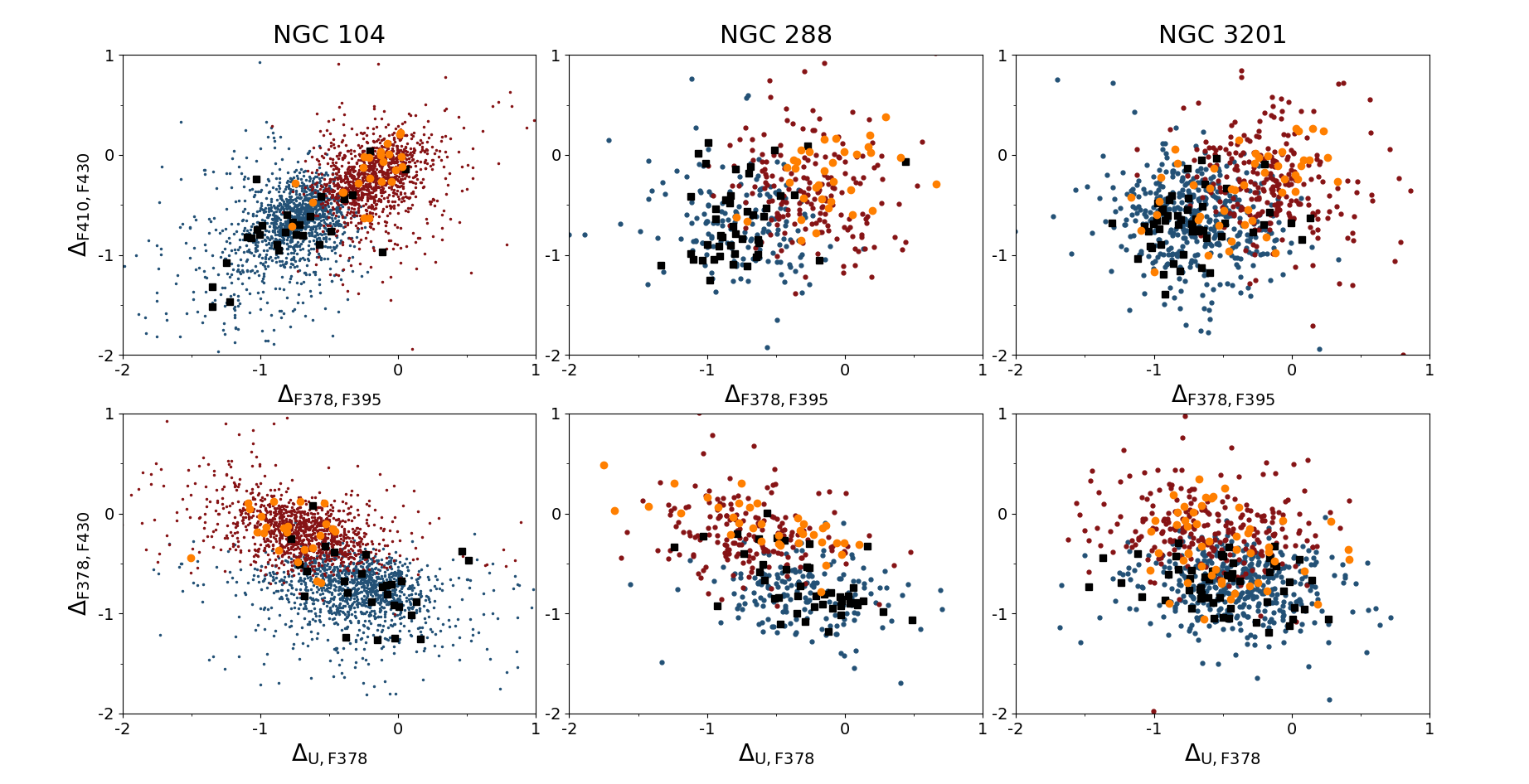}
    \caption{Two $\Delta$ combinations for three of the cluster in our sample. The blue and red points are the separation made using the clustering method. The black and orange points are stars with measured Na abundance from \citet{Carretta2009}. For NGC\,104 and NGC\,288, we can see the segregation of Na-poor (orange circles) and Na-rich (black squares) follows our separation. For NGC\,3201, the picture is less clear, and the Na-rich stars match very well with the blue population; however, the Na-rich do not occupy a clear locus.}
    \label{fig:Del_Abun}
\end{figure*}

To make full use of the six colour combinations selected, we used the K-means clustering algorithm \citep{Macqueen1967} with two classes to separate the populations present in the clusters. This method requires two main assumptions: first, that two distinct populations are present, as expected for the clusters in our sample, and second that the number of objects is comparable in both populations, as also supported by the literature. The separation was performed in all four clusters, and some $\Delta$ combinations for three GCs are shown in Fig.~\ref{fig:Del_Abun}. NGC\,7089 is a particular case, \citet{Milone2015} identified in this cluster 7 distinct populations. In their analyses using HST images they identified three main groups (A, B and C) with very distinct abundance patterns. However, given the low number of objects in population C we do not account for it in our method and do no find it in a visual inspection of the CMDs. Also, we are not capable of separating the more nuanced subpopulations of groups A and B given their relative similarity. This might be one of the reasons why the separation for NGC\,7089 is not clear in either the ${\Delta}colours$ planes and the CMD shown in Fig.~\ref{fig:CMD_Abun}.


\section{Analysis and discussion}
\label{S:Analyses}

\subsection{Classifying the populations}
\label{S:classifying}

\begin{figure}
    \centering
    \includegraphics[width=0.45\textwidth]{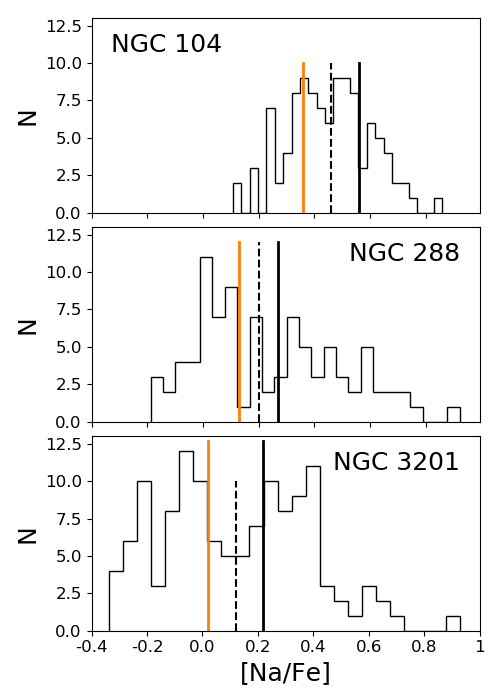}
    \caption{Sodium abundance histograms from \citet{Carretta2009} for three of the clusters. The Na-poor objects are left of the orange line, while the Na-rich is to the right of the black line.}
    \label{fig:Na_Hist}
\end{figure}

The usual understanding of the formation of MSPs states that a primordial population enriches ---by different methods depending on the model--- the intra-cluster medium forming a secondary population \citep{Decressin2007, D'Ercole2008, Charbonnel2014}. One of the elements that can be used to trace the two populations is Na, for which we have abundance measurements from \cite{Carretta2009} for individual stars in three of the GCs in our sample, namely NGC\,104, NGC\,288 and NGC\,3201. We separated the stars in Na-rich and Na-poor by first constructing the histogram of Na abundance is shown in Fig.~\ref{fig:Na_Hist} and determining the central dip in the number of objects (dashed line in Fig.~\ref{fig:Na_Hist}). We drew two more lines (orange and black) around this division taking into account the average uncertainty in [Na/Fe] with the intention of avoiding an ambiguous classification. Na-poor stars as those left of the orange line and Na-rich as those right of the black one. Last we matched with our S-PLUS photometry sample in order to evaluate our separation and to classify the populations. Figure \ref{fig:Del_Abun} shows two $\Delta$ combinations for each cluster with our separation as well as the Na-rich and Na-poor stars.

\begin{figure}
    \centering
    \includegraphics[width=0.5\textwidth]{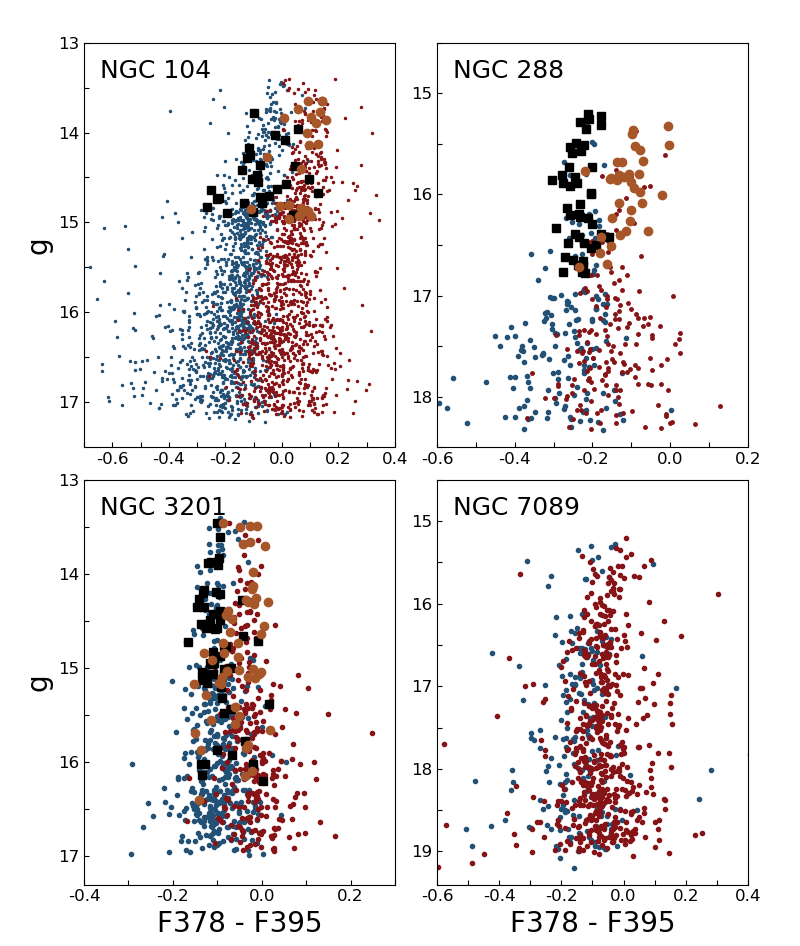}
    \caption{CMDs of only the RGB for the GCs in our sample. In blue and red are the populations separated using the K-means algorithm. The black squares and orange points are the same as in Fig.~\ref{fig:Del_Abun}. In the case of NGC\,3201, the separation between Na-rich/poor is evident for stars brighter than ~15.5 on the g band below this Na-poor stars seem not to follow the red population.}
    \label{fig:CMD_Abun}
\end{figure}

In general, we see that the objects with blue markers are connected with the Na enhanced stars. We consider this the second population (2P). The objects with a primordial composition (Na-poor) are connected with the objects in red, forming the first population (1P). For NGC\,104, this separation is very clear when looking at the top panel of Fig.~\ref{fig:Del_Abun}. NGC\,288 also presents a good separation in both $\Delta$ spaces. However, for NGC\,3201 the separation is fuzzy, with the Na-rich stars occupying a clearer locus matching with the blue population. Meanwhile, the Na-poor are more spread out over the entire distribution. Figure \ref{fig:CMD_Abun} shows one CMD for each cluster focused in the RGB with the populations separated as described and the Na-poor/rich stars.

The population ratios are shown in Tab.~\ref{tab:KS_AD}, NGC\,104 and NGC\,288 have an equal (47\% and 51\% of 1P, respectively) split between first and second population stars. For NGC\,3201, the second population dominates with 63\% of the cluster in a number of RGB stars. NGC\,7089 stands out as 73\% of its stars are in the first population.

\subsection{Radial Profiles}
\label{S:rdp}

The present-day distribution of the populations in a GC is a complex interplay between many factors such as the initial conditions (which are strongly dependent on the formation scenario) and the internal dynamical evolution of the cluster (mass-segregation, binaries, core-collapse, etc.) \citep{Vesperini2013, Vesperini2021, Calura2019, Sollima2021}. Looking at the cumulative distribution of the different populations as a function of their distance to the cluster centre is important when evaluating such formation scenarios; however, this must be done with some care. To check the significance of the radial difference between the distributions of the different stellar populations, we submit our radial profiles to a set of statistical tests. The most common one is the Kolmogorov-Smirnov test (KS-test) that evaluates whether two distributions differ significantly. However, the KS-test has some limitations, and it is less sensitive when the distributions differ in the beginning and end \citep{Feigelson2012}. Thus the Anderson-Darling test (AD-test) was designed to mitigate this \citep{Anderson1952}. Figure \ref{fig:RP_GCs} presents the radial profiles for the four clusters in our sample, and the radial distance is shown in terms of the half-light radius \citep[][2010 version]{Harris1996}. Table \ref{tab:KS_AD} shows the results for both tests as well as their critical values.

In the following subsections, we discuss the radial distributions of the MSPs of our sampled GCs in light of past literature studies.

\subsubsection{NGC\,104}

One of the most massive clusters in our galaxy, NGC\,104 is an interesting subject for our study. \citet{Norris_Freeman1979} measured the CN abundance in 142 RGB stars and found that the richer population is more centrally concentrated. This result was corroborated by \citet{Briley1997} who studied ~300 RGB stars and found that outside of 13 arcmins from the cluster centre, CN-weak stars dominate, while inside, no difference is apparent. \citet{Milone2012} used ground-based and HST photometry to study the presence of MSPs along the entire stellar sequence of NGC\,104. They found that the second population comprises $\sim$70\% of the cluster and is more centrally concentrated than the first up to 3-4 half-light radius. This result is in agreement with the work of \citet{Nataf2011} that studied the stars in the RGB bump and HB and found that the He-enhanced population is more centrally concentrated, however, this is a much more tenuous result. Looking at our results for NGC\,104, we can see that it is in agreement with the literature up to $\sim$3 half-light radii (9\,arcmin). Nonetheless, beyond this, no significant difference between the populations is apparent.

\subsubsection{NGC\,288}

\citet{Vanderbeke2015} studied the HB of 48 GCs and found that in the case of NGC\,288, NGC\,362 and NGC\,6218 the second population (He-rich) appears less centrally concentrated. However, this is in contradiction with what was found by \citet{Piotto2013}. They used HST imaging to show that inside the FoV of WFC3/UVIS, the first population makes up more than 50\% of the cluster stars. Our results show that the second population (Na-rich) is more centrally concentrated, supported by the KS and AD tests which show that the two distributions are different with a high confidence level, agreeing with \citeauthor{Piotto2013}

\subsubsection{NGC\,3201}

\begin{figure*}
    \centering
    \includegraphics[width=0.85\textwidth]{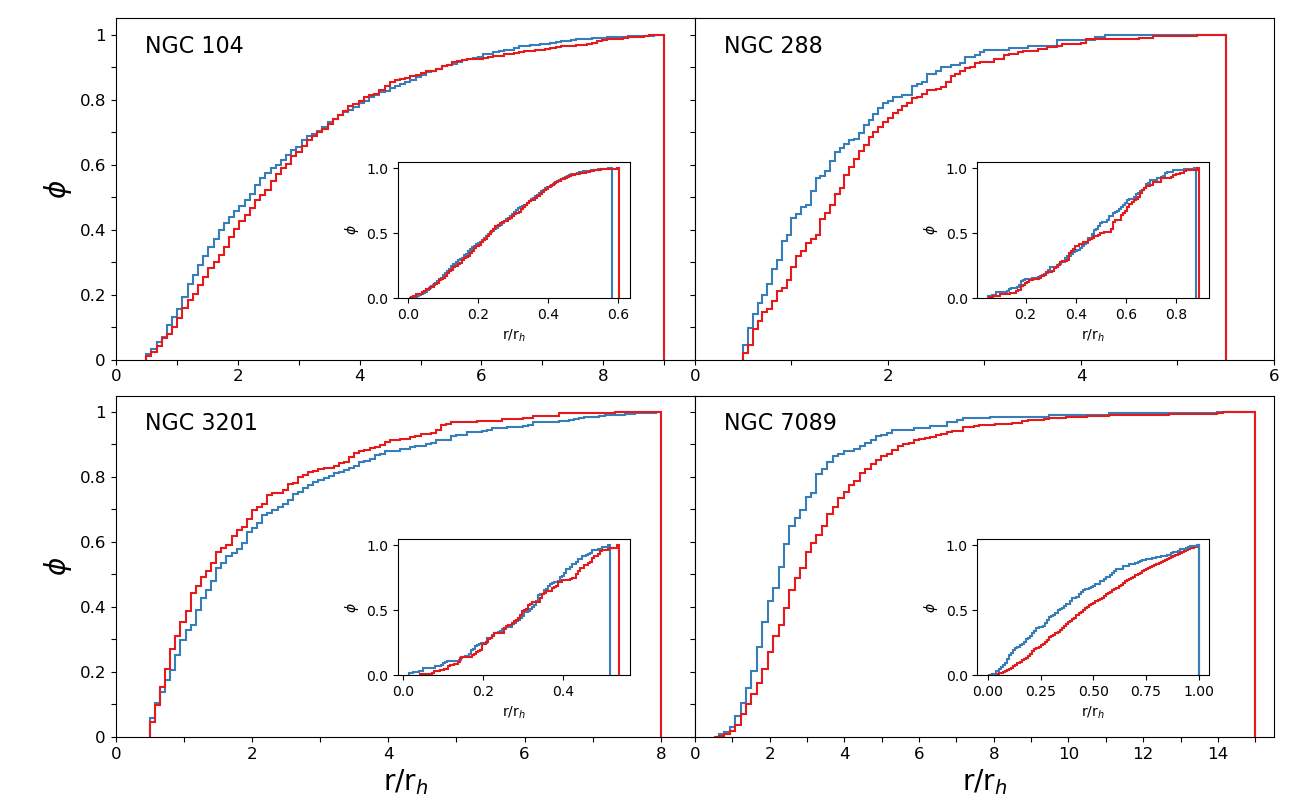}
    \caption{Cumulative radial distributions of the four clusters in our sample. The red and blue lines represent the primordial and enriched populations, respectively. The insets show the radial profiles for the inner region of the clusters constructed with HST photometry, highlighting the importance of studying the outer regions.}
    \label{fig:RP_GCs}
\end{figure*}

NGC\,3201 has been extensively studied by \citet{Kravtsov2010}, \citet{Kravtsov2017} and \citet{Kravtsov2021}. Overall they found with a high degree of confidence that the second population is more centrally concentrated. This result was consistent in the SGB and RGB and across different data sets. \citet{Carretta2010} also studied this cluster and corroborated these results, finding that Na-poor RGB stars occupy more the outskirts of the cluster, although their sample size was relatively small. The radial profile shown in Fig.~\ref{fig:RP_GCs} seems to contradict these results, showing a larger fraction of 1P stars towards the centre of the GC. This discrepancy may be due to the fact that the photometric separation provided by our calculated ${\Delta}colours$ does not provide a consistent separation between 1P (Na-poor) and 2P (Na-rich) as shown in Fig.~\ref{fig:Del_Abun}.

\subsubsection{NGC\,7089}

\citet{Vanderbeke2015} in the same study of HB stars mentioned earlier found no radial difference when analysing NGC\,7089 with a KS probability of 72\%. However, \citet{Lardo2011} used SDSS photometry and showed a large radial variation between both populations, finding that the UV-red (2P) population is more centrally concentrated. \citet{Hoogendam2021} have re-analysed the SDSS data and incorporated ground-based photometry by Stetson (2019) and found no conclusive evidence. The SDSS data set suggests a red-concentrated population, while Stetson's data shows the opposite, both to a high significance level in the KS test. According to the KS and AD tests, we find that the 2P of stars is more centrally concentrated with a high probability.

\subsection{The importance of looking at the outskirts}

Dynamical simulations of MSPs have shown that mixing occurs in a shorter time scale in the inner parts of the cluster due to two-body relaxation being more efficient in denser environments \citep{Vesperini2013}. Some information regarding the concentration of the second population, however, is still preserved for a longer period in the outer regions of the GCs. To test this and highlight the importance of wide-field studies of MSPs, we have taken HST archival data from the HUGS project \citep{Nardiello2018,Piotto2015}, reproduced the chromosome maps as described in \citet{Milone2012} and separated the populations according to \citet{Milone2017}. Given the smaller FoV of HST, the resulting radial profiles of all clusters extend at most to the half-light radius; they are in the insets in Fig. \ref{fig:RP_GCs}. In three of the clusters ---NGC\,104, NGC\,288 and NGC\,3201--- we see that the populations are already mixed, showing no major differences. The exception is NGC\,7089, where the populations show no sign of mixing with the second population appearing more centrally concentrated. This follows the trend found in the outer region of the cluster, and simulations by \citet{Dalessandro2019} suggest that in clusters with the dynamical age of NGC\,7089, some segregation can still be present inside 2 r$_{h}$.

\begin{table*}
\begin{tabular}{ccccccll}
\multicolumn{1}{l|}{} & \multicolumn{1}{l}{} & \multicolumn{1}{l|}{} & \multicolumn{1}{l|}{} & \multicolumn{2}{c|}{KS} & \multicolumn{2}{c}{AD}   \\
ID                    & N$_{blue}$           & N$_{red}$             & $f_1$                 & D        & P            & A     & A$_{cr}$         \\ \hline
NGC 104               & 1327                 & 1211                  & 0.47                  & 0.08     & 0.038\%      & 4.57  & 0.5\%            \\
NGC 288               & 209                  & 221                   & 0.51                  & 0.21     & 0.026\%      & 9.09  & \textless{}0.1\% \\
NGC 3201              & 494                  & 300                   & 0.37                  & 0.16     & 0.007\%      & 14.53 & \textless{}0.1\% \\
NGC 7089              & 199                  & 542                   & 0.73                  & 0.22     & 0.00019\%    & 16.68 & \textless{}0.1\%
\end{tabular}
\caption{\textbf{Population sizes and Statistics.} D and P: KS statistic and probability of the two distributions being drawn from the same parent population, A and A$_{cr}$: AD statistic and critical value.}
\label{tab:KS_AD}
\end{table*}


\section{Summary and Concluding Remarks}
\label{S:Conclusion}

When analysing the phenomenon of multiple populations, it is clear that the best tool is spectroscopy. It allows us to get a clear picture of the most significant differences between the populations and provide the best data set to inform possible formation models. However, it is an expensive tool that translates into studies with a relatively small sample size and limited to the outer regions of clusters. This is why photometric studies are so important, capable of providing information on thousands of objects at once. If we are capable of characterising the populations, it can provide us with a good picture of what is happening in the GCs. As we have shown here, S-PLUS is a great tool for this purpose. With its wide FoV, it is capable of studying the entirety of the cluster, and its set of 12 filters provides us with a large toolbox to analyse and separate the populations.

In the present study we have used six colour combinations (u - F378, u - F395, F378 - F395, F378 - F430, F378 - F515 and F410 - F430) and the K-means algorithm to separate the MSPs present in four GCs. We can see based on the selected filters that the spectral region that provides the best separation tends toward the blue, which is expected as it is in this region where the majority of MSP features appear. When combined with spectroscopic abundances of individual stars from the literature, our photometric separation is well supported in the cases of NGC\,104 and NGC\,288. However, when considering NGC\,3201, the separation does not seem to correspond to a difference in the Na abundance of the cluster stars. One thing to be noted here is that the more metal-rich GCs (NGC\,104 --- [Fe/H] = -0.72, NGC\,288 --- [Fe/H] = -1.32) in our sample have a more clear separation in the $\Delta$ space. This trend is in agreement with synthetic spectra computed to simulate MSPs (Branco et al., {\it in prep}). 

Using the large FoV of S-PLUS, we analysed the cumulative radial distribution (CRDs) of the populations. Using both the KS-test and the AD-test, we conclude that CRDs of the four clusters differ significantly. In the case of NGC\,104, the populations appear well mixed, which, given the age of this GC, could indicate that the populations had enough time to mix. For both NGC\,288 and NGC\,7089, we can see a clear concentration of 2P population toward the centre of the cluster. This directly supports the formation theories that propose an enrichment of the intra-cluster medium and subsequent star formation in the more dense central regions \citep[see e. g.][]{D'Ercole2008}. However, in the case of NGC\,3201, the trend is reversed. The 1P is more centrally concentrated, in direct contradiction with previous literature studies. It is clear that further studies have to be performed in a systematic way to shed light on this subject. Another critical issue is that in order to explain the differences in the clusters, formation scenarios have to be stochastic enough to account for the distinct histories of each GC.


\section*{Acknowledgements}
The authors thank the anonymous referee for the careful reading of the paper and the useful comments that helped improve the manuscript. This paper is based on the research of an undergraduate and masters thesis as part of the requirements for obtaining the title of Bachelor's and MSc in Physics at the \textit{Universidade Federal do Rio Grande do Sul}. This work was supported by Coordenação de Aperfeiçoamento de Pessoal de Nível Superior (CAPES), \textit {Conselho Nacional de Desenvolvimento Cient\'ifico e Tecnol\'ogico} (CNPq), \textit{Fundação de Amparo à Pesquisa do Estado do RS} (FAPERGS) and \textit{Fundação de Amparo à Pesquisa do Estado de SP} (FAPESP). The authors thank Analía Smith Castelli, Clecio R. Bom, Leandro Beraldo e Silva, Pavel Kroupa and Emanuele Dalessandro for comments that helped improve the manuscript. CB acknowledges funding from CNPq. 
ACS acknowledges funding from CNPq and FAPERGS through grants CNPq-403580/2016-1, CNPq-11153/2018-6, PqG/FAPERGS-17/2551-0001, FAPERGS/CAPES 19/2551-0000696-9, L'Or\'eal UNESCO ABC \emph{Para Mulheres na Ci\^encia} and the Chinese Academy of Sciences (CAS) President's International Fellowship Initiative (PIFI) through grant E085201009.
J.A.-G. acknowledges support from ANID – Millennium Science Initiative Program – ICN12\_009 awarded to the Millennium Institute of Astrophysics MAS.
This paper has made use of results from the European Space Agency (ESA) space mission Gaia, the data from which were processed by the Gaia Data Processing and Analysis Consortium (DPAC). Funding for the DPAC has been provided by national institutions, in particular the institutions participating in the Gaia Multilateral Agreement. The Gaia mission website is http: //www.cosmos.esa.int/gaia.
PC acknowledges support from Conselho Nacional de Desenvolvimento Cient\'ifico e Tecnol\'ogico (CNPq) under grant 310041/2018-0.

The S-PLUS project, including the T80-South robotic telescope and the S-PLUS scientific survey, was founded as a partnership between the Funda\c{c}\~{a}o de Amparo \`{a} Pesquisa do Estado de S\~{a}o Paulo (FAPESP), the Observat\'{o}rio Nacional (ON), the Federal University of Sergipe (UFS), and the Federal University of Santa Catarina (UFSC), with important financial and practical contributions from other collaborating institutes in Brazil, Chile (Universidad de La Serena), and Spain (Centro de Estudios de F\'{\i}sica del Cosmos de Arag\'{o}n, CEFCA). We further acknowledge financial support from the São Paulo Research Foundation (FAPESP), the Brazilian National Research Council (CNPq), the Coordination for the Improvement of Higher Education Personnel (CAPES), the Carlos Chagas Filho Rio de Janeiro State Research Foundation (FAPERJ), and the Brazilian Innovation Agency (FINEP).

The members of the S-PLUS collaboration are grateful for the contributions from CTIO staff in helping in the construction, commissioning and maintenance of the T80-South telescope and camera. We are also indebted to Rene Laporte, INPE, and Keith Taylor for their important contributions to the project. From CEFCA, we thank Antonio Mar\'{i}n-Franch for his invaluable contributions in the early phases of the project, David Crist{\'o}bal-Hornillos and his team for their help with the installation of the data reduction package \textsc{jype} version 0.9.9, C\'{e}sar \'{I}\~{n}iguez for providing 2D measurements of the filter transmissions, and all other staff members for their support with various aspects of the project. 
\section*{Data Availability}
The data underlying this article are available in its online supplementary material and at CDS via anonymous ftp to cdsarc.u-strasbg.fr (130.79.128.5) or at https://cdsarc.unistra.fr/viz-bin/cat/J/MNRAS and can be accessed with the volume and page numbers of this article.



\bibliographystyle{mnras}
\bibliography{Final} 



\appendix

\section{Field-Cluster Star Separation for the Other Clusters}
\label{app:selection}

\begin{figure*}
    \centering
    \includegraphics[width=0.85\textwidth]{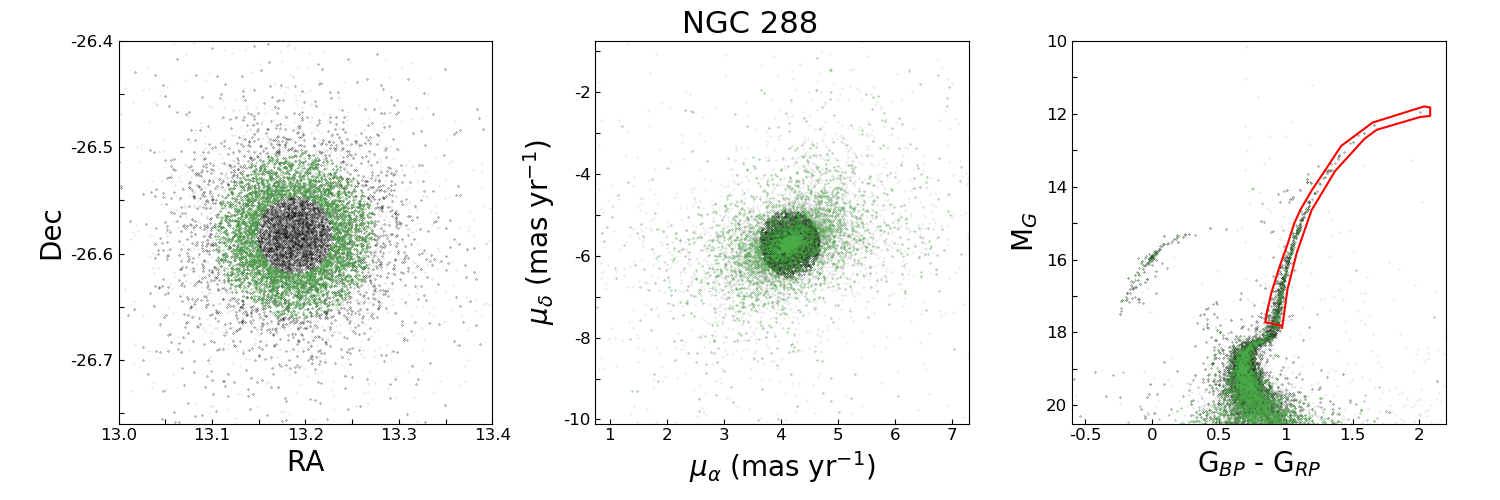}
    \caption{Same as Fig.~\ref{fig:GAIA_selection} for NGC\,288.}
    \label{fig:GAIA_NGC288}
\end{figure*}

\begin{figure*}
    \centering
    \includegraphics[width=0.85\textwidth]{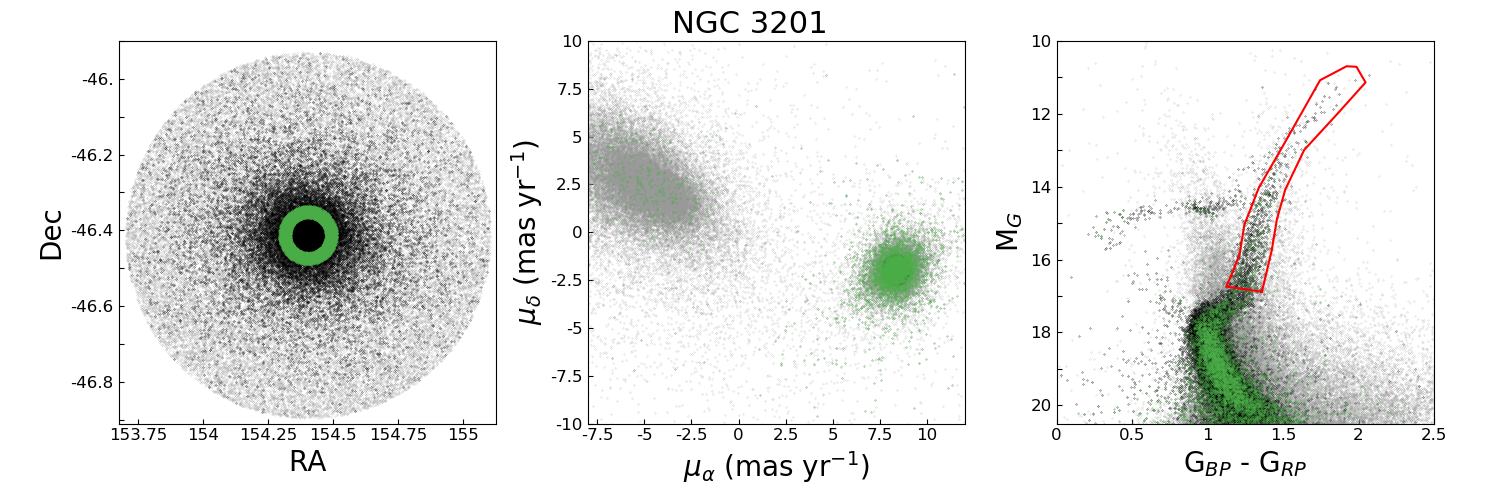}
    \caption{Same as Fig.~\ref{fig:GAIA_selection} for NGC\,3201.}
    \label{fig:GAIA_NGC3201}
\end{figure*}

\begin{figure*}
    \centering
    \includegraphics[width=0.85\textwidth]{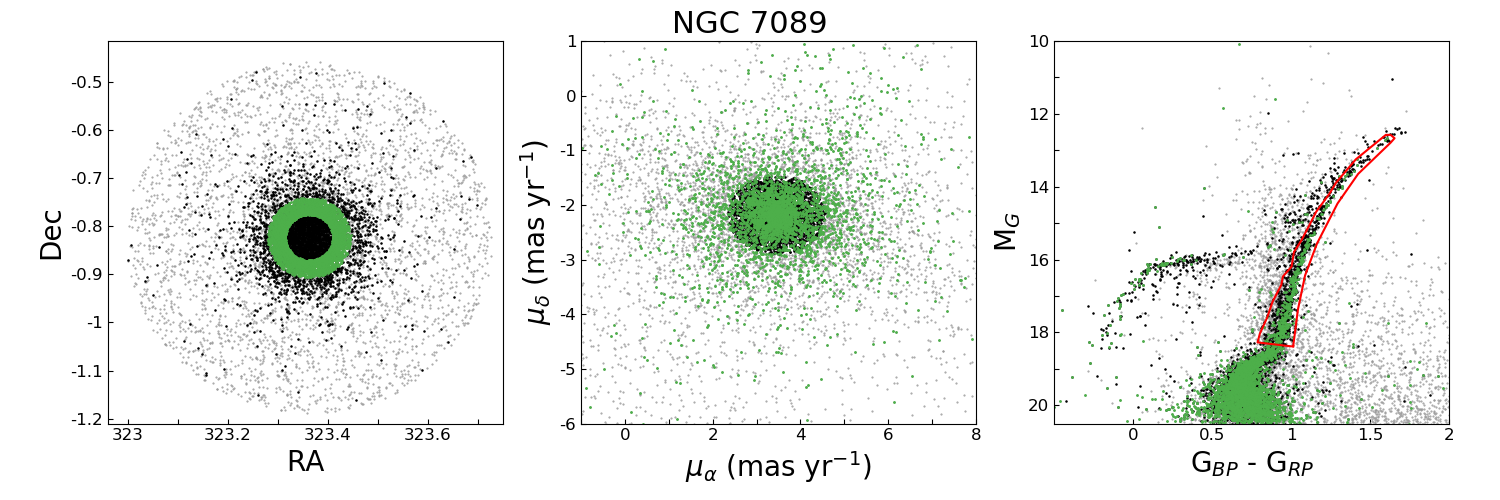}
    \caption{Same as Fig.~\ref{fig:GAIA_selection} for NGC\,7089.}
    \label{fig:GAIA_NGC7089}
\end{figure*}

In Fig.\, \ref{fig:GAIA_NGC288}, \ref{fig:GAIA_NGC3201} and \ref{fig:GAIA_NGC7089}  we show the process of cluster member selection using GAIA proper motions, as outlined in section \ref{S:selection}, for clusters NGC\,288, NGC\,3201 and NGC\,7089 respectively. 

\section{Incompleteness in the GAIA sample}

To analyse the completeness of the GAIA sample we constructed histograms of the number of objects as a function of the M$_{G}$ magnitude. We also marked the approximate position of the turn-off for each cluster. It is evident that the completeness is essentially unaffected above the TO.

\begin{figure*}
    \centering
    \includegraphics[width=0.95\textwidth]{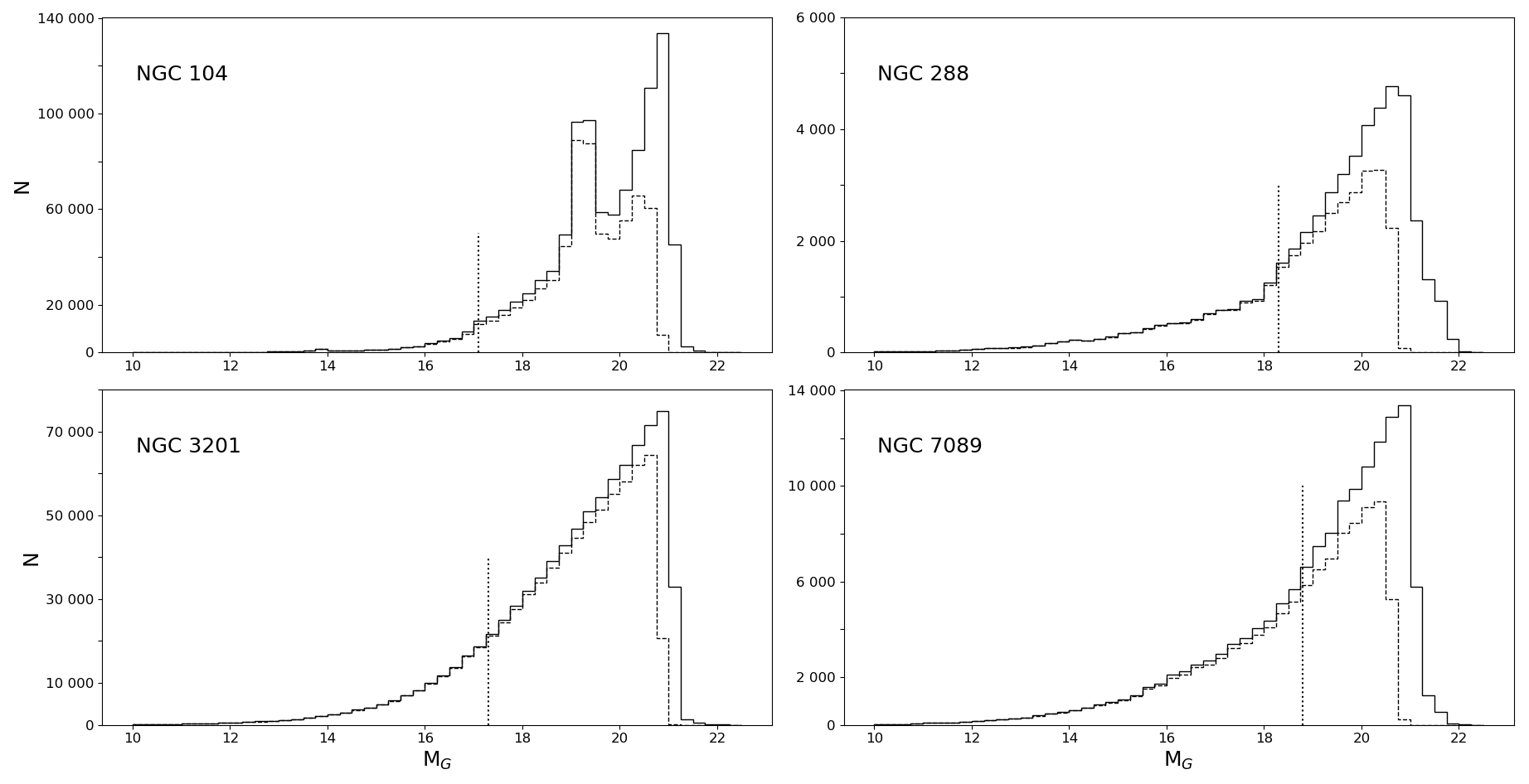}
    \caption{Histograms of the number of objects per magnitude bin. The solid line represent all the objects present within the tidal radius of each cluster, the dashed one are only those that have all three GAIA magnitudes measured and proper motion errors smaller than 1.5 mas yr$^{-1}$. The dotted line is the approximate position of the turn-off.}
    \label{fig:Completness}
\end{figure*}


\bsp	
\label{lastpage}
\end{document}